\documentclass[pra,reprint]{revtex4-1}
\usepackage[normalem]{ulem}
\usepackage{graphicx}
\usepackage{hyperref}
\usepackage{amsmath}
\usepackage{mathtools}
\usepackage{bm}
\usepackage{bbold}
\DeclareMathOperator{\sgn}{sgn}

\begin{document}
\title{Chern numbers for the index surfaces of photonic crystals: \\ conical refraction as a basis for topological materials}
\author{R. L. Mc Guinness}
\author{P. R. Eastham}
\affiliation{School of Physics and CRANN, Trinity College Dublin, Dublin 2, Ireland.}

\date{\today}

\begin{abstract}
The classification of bandstructures by topological invariants
provides a powerful tool for understanding phenomena such as the
quantum Hall effect. This classification was originally developed in
the context of electrons, but can also be applied to photonic
crystals. In this paper we study the topological classification of the
refractive index surfaces of two-dimensional photonic crystals. We
consider crystals formed from birefringent materials, in which the
constitutive relation provides an optical spin-orbit coupling.  We
show that this coupling, in conjunction with optical activity, can
lead to a gapped set of index surfaces with non-zero Chern
numbers. This method for designing photonic Chern insulators exploits
birefringence rather than lattice structure, and does not require band
crossings originating from specific lattice geometries.
\end{abstract}


\maketitle

\section{Introduction}

Chern insulators are two-dimensional materials that are insulating in
their interior but conducting along their edge. The first such
materials to be identified were the integer quantum Hall states
\cite{KlTZ}, for which the quantized Hall conductance corresponds to a
topological invariant known as the Chern
number\ \cite{AvSeiSi,KOHM,TKNN}. The Chern number is related to the
number of conductive edge states at the interface between a
topologically non-trivial insulator and a topologically trivial one
\cite{Hats}. A later example of a Chern insulator is provided by the
Haldane model \cite{HLDN}, which introduced a now well-trodden path to
constructing topologically non-trivial bandstructures. This model
describes particles hopping on a two-dimensional hexagonal lattice, an
arrangement which produces point degeneracies in the
bandstructure. These are the celebrated Dirac points, which are
associated with vortex-like singularities in the Bloch functions. They
can be a precursor to a topologically non-trivial gapped
bandstructure, which arises if the degeneracies are split by a
perturbation in such a way that the windings of the vortices combine
rather than cancel \cite{SParis}.

Topologically non-trivial bandstructures are not restricted to
theories of electrons, but can also be achieved for photons
\cite{HLDRAG,HLDRAG2,PGRPHN,Wang1,Wang2,PFloq,SelfGuidedEdge,PTI,Chen,BMWPTI,3DPTI},
polaritons \cite{PolZG,PolZ,TopPol,YiKar}, and sound waves
\cite{TopAc-1,TopAc0,TopAc1}.  The possibility of topological photonic
bands was raised by Haldane and Raghu \cite{HLDRAG,HLDRAG2}, who
showed how the original Haldane model could be implemented for light
in a hexagonal photonic crystal. Many subsequent works have followed
this proposal, showing how the degeneracies of photonic lattices can
be split to achieve either a $Z$ photonic Chern insulator
\cite{HLDRAG2,PGRPHN,PFloq} or a $Z_2$ photonic topological insulator
\cite{PTI,Chen}. Often such works consider situations where the
polarization and propagation decouple as, for example, for the TE and
TM modes propagating in the plane of a 2D photonic crystal. However,
such decoupling does not occur in an optically anisotropic material or
structure, where the polarization states depend on the wavevector
direction.

The topological classification of photonic materials is usually based
on their dispersion relation, i.e., the bandstructure, and the
associated Bloch states. Another important quantity, however, is the
refractive index surface. This is related to, but distinct from, the
dispersion relation: it is the surface of wavevectors corresponding to
a particular frequency. In this paper we study topological effects
that derive from the refractive index surface. We show how the
generic features of the index surfaces of anisotropic bulk materials
lead, in periodic systems based on such materials, to a type of
photonic Chern insulator.

\begin{figure}[!htb]
\includegraphics[width=\linewidth]{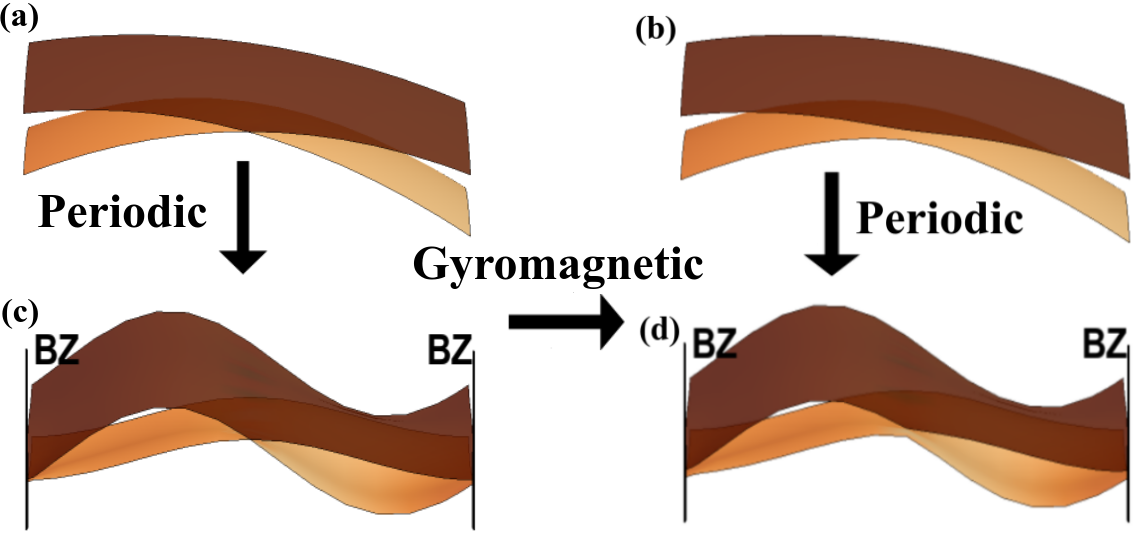}
\caption{(Color online) Illustration of the formation of topologically non-trivial
  index surfaces in two-dimensional photonic crystals. Each panel is a
  schematic of a section of a refractive index (isofrequency) surface,
  for (a)a biaxial dielectric, (b)a biaxial dielectric with optical
  activity, (c)a periodic biaxial dielectric, (d)a periodic biaxial
  dielectric with optical activity. Each panel is centered on the
  wavevector corresponding to the conical singularity in (a). The
  periodicity is taken to be in the plane perpendicular to this
  wavevector, forming a two-dimensional Brillouin zone
  (BZ).\label{fig:cartoon}}
\end{figure}

Our focus is on two-dimensional photonic crystals formed from biaxial
materials or metamaterials. For such materials the index surface
consists of two ellipsoids that intersect at conical singularities,
which are point degeneracies that are equivalent to the Dirac points
of a dispersion relation. In a periodic system the index surface
consists of many sheets defined on a two-dimensional Brillouin zone,
and contains conical intersections inherited from those of the bulk,
as shown in Fig.\ \ref{fig:cartoon}. These can be split in such a way as to
achieve a gapped set of index surfaces with non-zero Chern numbers,
i.e., a form of optical Chern insulator. Our approach is based on
effective medium theory, and exploits the singularities that are
generic, topologically enforced features of optical index surfaces,
rather than those of particular lattices. This suggests that similar
Chern index surfaces could be achieved in two-dimensional photonic
structures with a range of lattice geometries.

A related concept of photonic Chern insulators has been put forward by
Gao et al.~\cite{Gao}. That work, however, considers an optically
homogeneous material, for which the refractive index is defined on the
sphere of wavevector directions. We consider instead a periodic
material, and specifically a two-dimensional photonic crystal. In this
case the refractive index is defined not on a sphere, but on a
two-dimensional torus, i.e. the Brillouin zone. This change in the
underlying topology necessitates a different approach to designing
topological materials. We note also that the topological
classification of the index surface of a periodic structure is
implicit in the tight-binding model of the photonic Floquet
topological insulator, as realized on a hexagonal
lattice~\cite{PFloq}. The present work places this type of state
within the broader context of the optics of periodic anisotropic
materials, and reveals a different approach to its realization.

\section{Model}

We consider the propagation of light, at some fixed frequency
$\omega$, through a two-dimensional photonic crystal, and seek to map
this problem to a Schrodinger equation with a topologically
non-trivial Hamiltonian. We do this via the refractive index surface,
which is a polar plot of the refractive index $n$ over all possible
wavevector directions. Since $n=ck/\omega$, where $k$ is the magnitude
of the wavevector, the index surface is a constant-frequency surface
in wavevector space, akin to the Fermi surface of a solid. The
connection to a Schrodinger equation follows on noting that the index
surface determines one wavevector component (propagation constant) in
terms of the other two, $k_z=\sqrt{n^2 \omega^2/c^2-k_x^2-k_y^2}$, so
for a scalar field we have $i\partial_z
\psi(k_x,k_y,z)=k_z\psi(k_x,k_y,z)\equiv H\psi(k_x,k_y,z)$. The
polarization of light may be included by replacing $\psi$ with a
two-component field, formed from the complex amplitudes of two
orthogonal polarization states. In an anisotropic material there will
be two distinct refractive indices $n_{\pm}$ for each wavevector
direction, each associated with a particular, wavevector-dependent
polarization. Thus $H$ becomes an operator acting in both spin
(polarization) space and real space, with spin-orbit coupling
terms~\cite{Bry3}.

The forms of index surface for various dielectrics are well
known~\cite{bornandwolf,landau_electrodynamics_2008}. We consider here
the most general case of a biaxial material. For such materials the
dielectric tensor possesses three distinct eigenvalues $\epsilon_1\neq
\epsilon_2 \neq \epsilon_3$, and the index surface consists of two
ellipsoids. These ellipsoids intersect at four conical singularities,
one of which is shown in Fig. 1a. The wavevectors of these
singularities are the optic axes of the crystal, along which light
undergoes conical refraction \cite{Bry2}. Near a singularity the
effective Hamiltonian, in the basis of circular polarization states,
takes the Dirac-point form $H=A\hat{\bm{\sigma}}.(k_x,k_y,0)$, where
$k_x,k_y$ are components of the wavevector perpendicular to the optic
axis \ \cite{Bry3}. The quantity $A$, which plays the role of the
Fermi velocity, is the semi-angle of the conical intersection. If we
introduce a periodic modulation in a plane containing the wavevectors
$k_x$ and $k_y$ then they become defined on a two-dimensional
Brillouin zone, and additional singularities appear in the index
surface (Fig. 1c). Optical activity splits these singularities
(Figs. 1b, 1d) and, as we show in the following, this can be done in
such a way as to achieve a non-zero Chern number.

\subsection{Effective Hamiltonian: paraxial approximation}

To demonstrate the formation of non-zero Chern numbers for the index
surface we will consider the specific case of a two-dimensional
photonic crystal, with periodicity perpendicular to an optic axis. We
will develop an effective Hamiltonian based on the paraxial
approximation to the index surface of a homogeneous dielectric. As
discussed in Refs.\ \onlinecite{Bry1,Bry2,Bry3,Bry5,Jfry}, considering
plane-wave solutions to Maxwell's equations in such a dielectric leads
to an eigenvalue problem
\begin{equation}
Md_{\pm}=\lambda~d_{\pm}=d_{\pm}/n_{\pm}^2,\label{eq:genindxsurf}
\end{equation} where the two-dimensional matrix $M$ depends on the propagation direction $\hat{\mathbf{k}}$ and the dielectric tensor. For each direction there are two distinct refractive indices $n_{\pm}$, and two corresponding polarization states. Each of the latter is specified by the two-component complex vectors $d_{\pm}$ for which we use the basis of circular polarizations.

As discussed above, Eq.  (\ref{eq:genindxsurf}) corresponds to a
two-dimensional Hamiltonian for propagating the field forwards in $z$;
this Hamiltonian is a matrix whose eigenvalues are
$k_{z,\pm}(k_x,k_y)$, and whose eigenvectors are
$d_{\pm}(k_x,k_y)$. Since the behavior of the homogeneous dielectric
is scale-invariant we may introduce the characteristic wavevector
$k_0=\omega/c$ and write $i\partial_z\psi=k_0 H\psi$, where $H$ is the
matrix whose eigenvalues are
$\sqrt{n_{\pm}^2-\tilde{k}_x^2-\tilde{k}_y^2}$, where $\tilde{k}_x=k_x/k_0$. We now
specialize to consider wavevectors which are close to the z direction.
Thus $\tilde{k}_x,\tilde{k}_y\ll 1$ and we may expand $H$ as a power
series in these quantities. This leads to a paraxial
approximation \begin{equation} H_{p}=h_0(\tilde{k}_x,\tilde{k}_y) I +
  \bm{h}(\tilde{k}_x,\tilde{k}_y).\hat{\bm{\sigma}},\label{eq:Ham}\end{equation}
where $h_0$ and $\bm{h}$ are second-order polynomials in the off-axis
components of wave vector. The forms of these
polynomials are given in the appendix, for the case of propagation
near to an optic axis in a material with biaxiality and anisotropic
optical activity.

The paraxial Hamiltonian $H_{p}$ gives a local approximation to the
index surfaces near the propagation direction, taken to lie along the
optic axis. This local approximation should give a reasonable account
of the topological structure of the index surface so long as the range
of transverse wavevectors does not encompass any other
singularities. The part of the Hamiltonian proportional to
$h_0$ controls the overall curvature of the index surfaces. In the
absence of optical activity, the index surface is split into two
linearly-polarized surfaces due to the off-diagonal components
proportional to $h_x$ and $h_y$, which vanish at the optic
axis. Optical activity lifts the degeneracy of the two circular
polarizations at these points, and the index surfaces are gapped
everywhere. As we shall see, a non-zero Chern number in the periodic
case requires an anisotropic optical activity. For definiteness we
suppose this to arise from the Faraday effect, which is parameterized
by an optical activity vector $\bm{g}$ that is related to the applied
field (see appendix).

\subsection{Periodic generalization}

The Chern numbers of a bandstructure are integer topological
invariants, which characterize the mapping between the closed
two-dimensional Brillouin zone and the states defined by the 
Hamiltonian\ \cite{AvSeiSi,TKNN}. The states for each band
$|\psi(k_x,k_y)\rangle$ provide a $U(1)$ Berry connection
$A_{i}=i\langle \psi|\partial_{k_i}|\psi\rangle$ and flux
$F=\partial_xA_y-\partial_yA_x$ , whose integral over the Brillouin
zone is $2\pi$ times the Chern number of the
band\ \cite{fukui_chern_2005}. The Hamiltonian $H_p$ discussed above,
however, does not define a quantized Chern number, because it is
defined on an open disk of wavevectors $k_x^2+k_y^2<n_-^2k_0^2$ rather than a
closed, two-dimensional Brillouin zone. We must therefore generalize
it so as to apply to a periodic structure. 

In order to do this we will consider the lattice version of the
Hamiltonian $H_p$, for simplicity considering a square lattice in the
$x-y$ plane with lattice constant $a$. The real-space lattice
Hamiltonian is obtained from $H_p$ by replacing the derivatives,
$k_{i}=-i\partial_{i}$, with finite differences. In wavevector space
this corresponds to making the replacements
\begin{equation} 
\begin{split}
&k_i\rightarrow\frac{1}{a} \sin k_ia\\
&k_i^2\rightarrow\frac{2}{a^2}[1- \cos k_ia],
\end{split}\label{eq:periodicity}
\end{equation} giving the wavevectors, and hence the Hamiltonian, the appropriate periodicity. As we shall see, this periodic Hamiltonian gives a qualitative description of the topological features of the index surfaces obtained from plane-wave calculations. It it is equivalent to a tight-binding model at a particular choice of hopping parameters. We shall
denote it by $H_{l}$ and introduce the wavevector scale in the
material $k=\sqrt{\epsilon_2}k_0$ for later convenience.

The effect of introducing periodicity is illustrated in
Fig.\ \ref{fig:cartoon}, which shows the evolution of the index
surfaces of a homogeneous biaxial material as optical activity,
periodicity, or both are added. The key feature of the reformulated
Hamiltonian $H_{l}$ is that, in the absence of optical activity, there
are additional degeneracies in the first Brillouin zone. This is a
consequence of the periodic topology of the Brillouin zone, which
requires the vector field $(h_{x},h_{y},h_{z}=0)$ to have zero net
circulation.  We shall now move to classifying the topological
character of the index surfaces of these photonic crystals.\\

\section{Results}

\subsection{Topological phases of the lattice model}

The topological phases of the lattice model can be deduced from the
degeneracy structure of the Hamiltonian $H_l$\ \cite{SParis}. For a
two-band Hamiltonian such as $H_l$ the Chern number, $C$, is the
winding number of the vector field $\bm{h}(k_x,k_y)$. It can be
computed by summing over the zeros of $(h_x,h_y)$ in the Brillouin
zone \begin{equation} C=\sum_i v_i
  \sgn h_z(i),\label{eq:chernfromh}\end{equation} where $i$ denotes a
zero of $(h_x,h_y)$ in the Brillouin zone, at which the vorticity is
$v_i$ and the mass term $h_z(i)$.

\begin{figure}[!htb]
\includegraphics[scale=0.4]{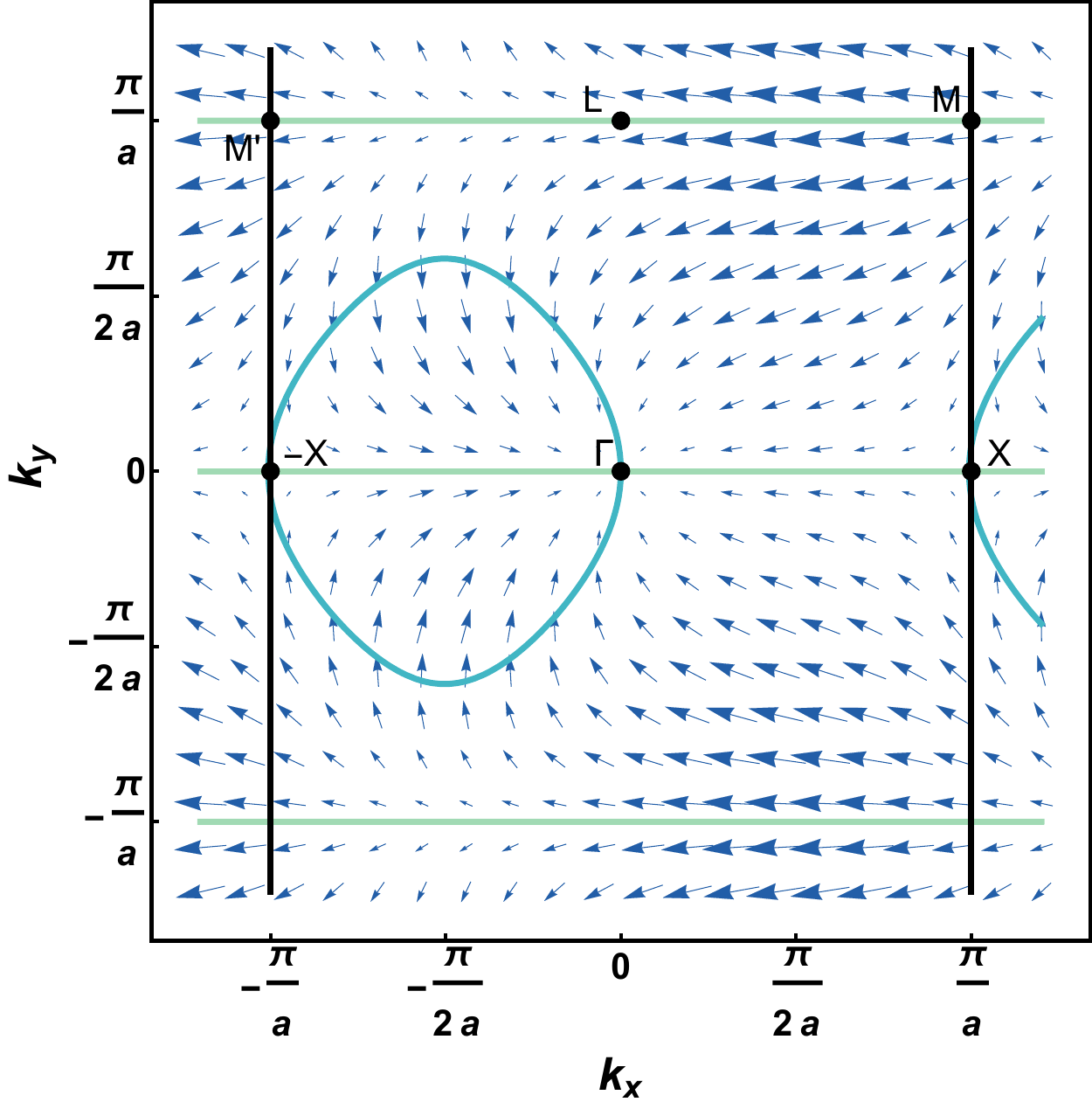}
\caption{(Color online) Polarization structure of the index surface over the first
  Brillouin zone for a biaxial material and a square lattice, predicted by Eqs. (\ref{eq:Ham}) and (\ref{eq:periodicity}). The arrows show the field $(h_x,h_y)$. The zeros of $h_x$ and $h_y$ lie along the lines indicated in turquoise (dark grey) and green (light grey), respectively. In the absence of optical activity $h_z=0$ and there are conical singularities at the intersections of the contours shown. The first Brillouin zone is the square bounded by the two black lines and the upper and lower green contours. The parameters are $\epsilon_1=2.25$, $\epsilon_2=2.5$, $\epsilon_3=2.75$ and $ka=1.6\pi$, where $k=\omega/(c\sqrt{\epsilon_2})$. \label{Fig2}  }  
\end{figure}

We therefore discuss first the case of zero magnetic field, so that
$h_z=0$. Fig.\ \ref{Fig2} shows the field
$(h_x,h_y)=|h|(\cos 2\theta,\sin 2\theta)$, along with the zero
contour lines of these components.  This field shows the direction of
linear polarization $\theta$ for the outermost sheet of the index
surface. The intersections of the contour lines are the conical
singularities or Dirac points, where the sheets of the index surface
are degenerate. 
For illustration we take the permittivities $\epsilon_1=2.25$,
$\epsilon_2=2.5$, $\epsilon_3=2.75$ and a lattice spacing to
wavelength ratio of $a/\lambda=4/5$. We use these
permittivities for the remainder of this paper, and this lattice
spacing to wavelength ratio except where otherwise stated.

As can be seen in Fig.\ \ref{Fig2}, the lattice periodicity has
introduced additional conical singularities in the index
surfaces. These are at the intersections of the zero contour lines of
$h_x$ (turquoise) and $h_y$ (green), and for these parameters there are
two such intersections in the first Brillouin zone. In addition to the
Dirac point at $\Gamma$, corresponding to that of the original index
surface, there is a second Dirac point just inside the $X$ boundary
along the line $k_y=0$. The Berry flux corresponding to each of these
singularities is $\pm\pi$, so that with two degeneracies in the first
Brillouin zone we can achieve a Chern number $C \in \{-1,0,1\}$.

Optical activity due to the Faraday effect introduces a term $h_z\neq
0$ and lifts the degeneracies in the index surface. The polarization
over the index surface then becomes elliptical, with a circularity
determined by the sign of $h_z$. To achieve a non-zero Chern number
the zero contour of $h_z(k_x,k_y)$ must separate the two lifted
degeneracies, so that $h_z$ has opposite signs at each of them, and
the signs of their Berry fluxes are the same
[Eq.\ (\ref{eq:chernfromh})]. Physically, the circular polarization
must swap between the two C-points in each sheet around which the
linear polarization winds. Whether or not this occurs depends on the
direction of the optical activity vector.

\begin{figure}[!htb]
\centering
\includegraphics[scale=0.4]{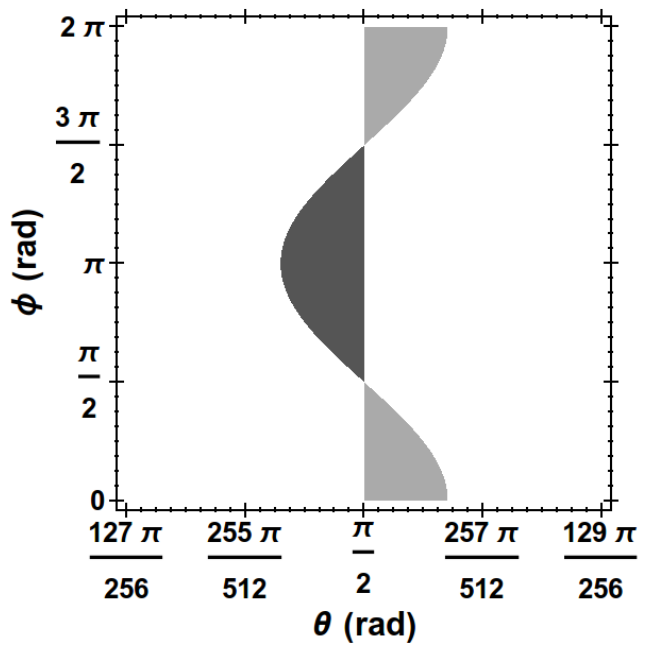}
\caption{Chern number $C$ (shading) of a biaxial material on a square
  lattice, from the paraxial Hamiltonian model, with optical activity
  due to the Faraday effect. The spherical coordinates $(\theta,
  \phi)$ give the direction of the axis of optical activity, i.e. the
  magnetic field. The polar angle $\theta$ is measured from the optic
  axis, and the azimuthal angle $\phi$ from the x-axis of the rotated
  coordinate system described in the appendix.  The shadings represent
  $C=0$ (white) and $C=\pm 1$ (dark/light gray). The parameters are
  those given in Fig.\ \ref{Fig2}.\label{Fig3}}
\end{figure}

In Figure \ref{Fig3} we show how the Chern number depends on the
direction of the optical activity vector. It shows the Chern number by
shading for a range of directions $(\theta,\phi)$ about and around the
equator $\theta=\frac{\pi}{2}$ at which the field and optic axis are
perpendicular.  We see that non-zero Chern numbers (light and dark
shading) can be achieved, for this model, when the optical activity
vector lies almost perpendicular to the optic axis. This region is
bounded by two equators: one perpendicular to the optic axis, which
corresponds to gap closing at the $\Gamma$ point, and one at a
slightly different angle, which corresponds to the gap closing at the
C-point near to $X$. The existence of a finite region between these
two equators, where the Chern number is non-zero, is due, primarily,
to the displacement of this latter C-point from the zone boundary.

The phase diagram in Figure\ \ref{Fig3} does not depend on the
strength of the optical activity, although of course this
parameter does affect the size of the gap. The dependence we have
observed on the remaining parameters of the model is related to the
displacement of the C-point away from the zone boundary. In the model
$H_l$ this displacement arises from the interplay of the linear and
quadratic terms in the paraxial Hamiltonian, which give different
contributions to the lattice model that vanish at different points in
the zone [see Eq. (\ref{eq:periodicity})]. If we increase the cone
angle $A$, i.e. the degree of biaxiality, the C-point near $X$ moves
into the Brillouin zone along the $k_y=0$ line, and the region of
non-zero Chern number in Fig.\ \ref{Fig3} expands. We can also vary
the ratio of the lattice spacing to the operating wavelength, $ka$,
which controls the paraxial approximation, Eq. (\ref{eq:Ham}). As this
parameter increases the second-order terms decrease relative to the
first-order ones, bringing the C-point near $X$ closer to the zone
boundary, and reducing the region of non-zero Chern number.

\subsection{Simulated index surfaces}

In this subsection we compare the periodic Hamiltonian model, $H_l$,
to the index surfaces extracted from a plane-wave calculation of a
two-dimensional photonic crystal\ \cite{MPB}. We focus on the
locations of the singularities in a material without optical activity,
since this is the key feature determining the topological phase
diagram when optical activity is introduced. We consider specifically
a biaxial dielectric in which cylindrical air holes are drilled to
form a square lattice.

\begin{figure}[!htb]
  \begin{minipage}{0.45\textwidth}
    \includegraphics[width=0.8\textwidth]{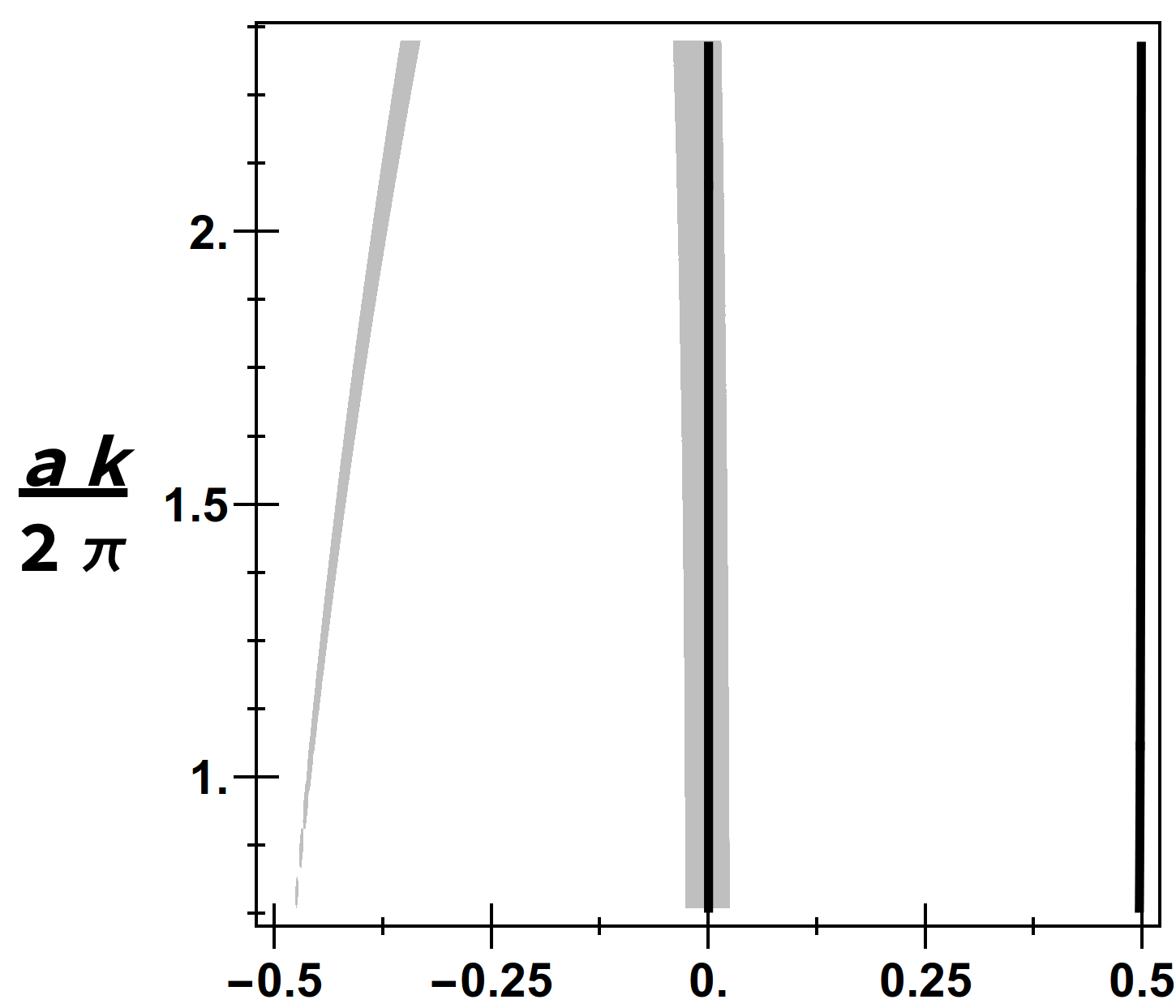}
  \end{minipage}
 \begin{minipage}{0.45\textwidth}
    \includegraphics[width=0.8\textwidth]{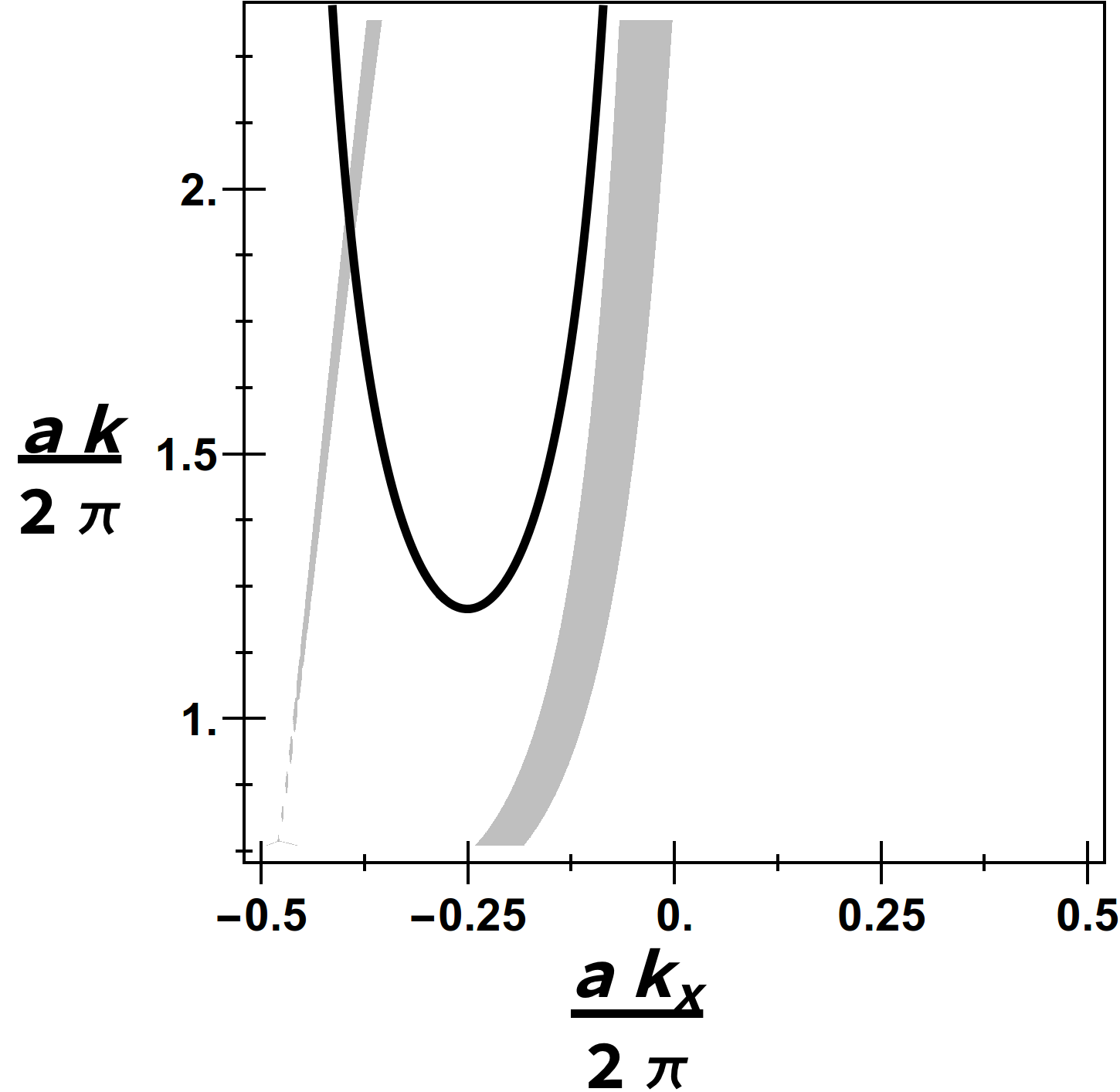}
  \end{minipage}
\caption{Comparison of the locations of the Dirac points given by the
  paraxial theory and a frequency-domain plane-wave simulation as
  $k$ is varied. In both cases Dirac
  points occur along the lines $k_y=0$ (top) and $k_y=\pi/a$
  (bottom). Solid lines: $k_x$ for the Dirac points from the paraxial
  theory. Shading: bounds on $k_x$ for the Dirac points from the
  plane-wave simulations, where the magnitude of the splitting between
  the index surfaces is smaller than $a\Delta k_z=2\pi \times
  e^{-6}$. The dielectric parameters are those in
  Fig.\ \ref{Fig2}. The plane-wave simulation has cylindrical air
  holes in the dielectric background, with a filling factor of
  $\pi(0.15)^2$. \label{Fig4}}
\end{figure}

In Fig.\ \ref{Fig4} we compare how the locations of the Dirac points
evolve with the wavevector scale $k$, at fixed $a$, in the two
theories. The solid black lines show the locations predicted by $H_l$,
while those predicted numerically lie within the shaded regions of the
figure. We see that $H_l$ gives a reasonable account of the numerical
results. As shown in the upper panel, it correctly predicts the two
Dirac points along the line $k_y=0$ for all values of $ka$, i.e., for
all degrees of paraxiality. As expected there is a Dirac point along
$k_y=0$ at $k_x=0$ in both theories. There is also a second Dirac
point present in each theory along this line. For the periodic
Hamiltonian model the second Dirac point is adjacent to the positive
$k_x$ Brillouin zone boundary; for the the numerical simulations the
Dirac point is found on the other side of this boundary displaced from
the zone edge. We attribute this difference to the different treatment
of scattering in the two methods, and perhaps also to higher-order
terms neglected in Eq. (\ref{eq:Ham}). The lower panel of
Fig. \ref{Fig4} shows the positions of Dirac points along the line
$k_y=\pi/a$.  While for small $ka$ the model $H_l$ lacks the two Dirac
points seen in the numerical simulation, there is a critical $ka$
above which this additional pair emerges. Above this critical $ka$ the
model $H_l$ is in qualitative agreement with the numerical simulation.

\begin{figure}[!htb]
  \begin{minipage}{0.45\textwidth}
    \includegraphics[width=\textwidth]{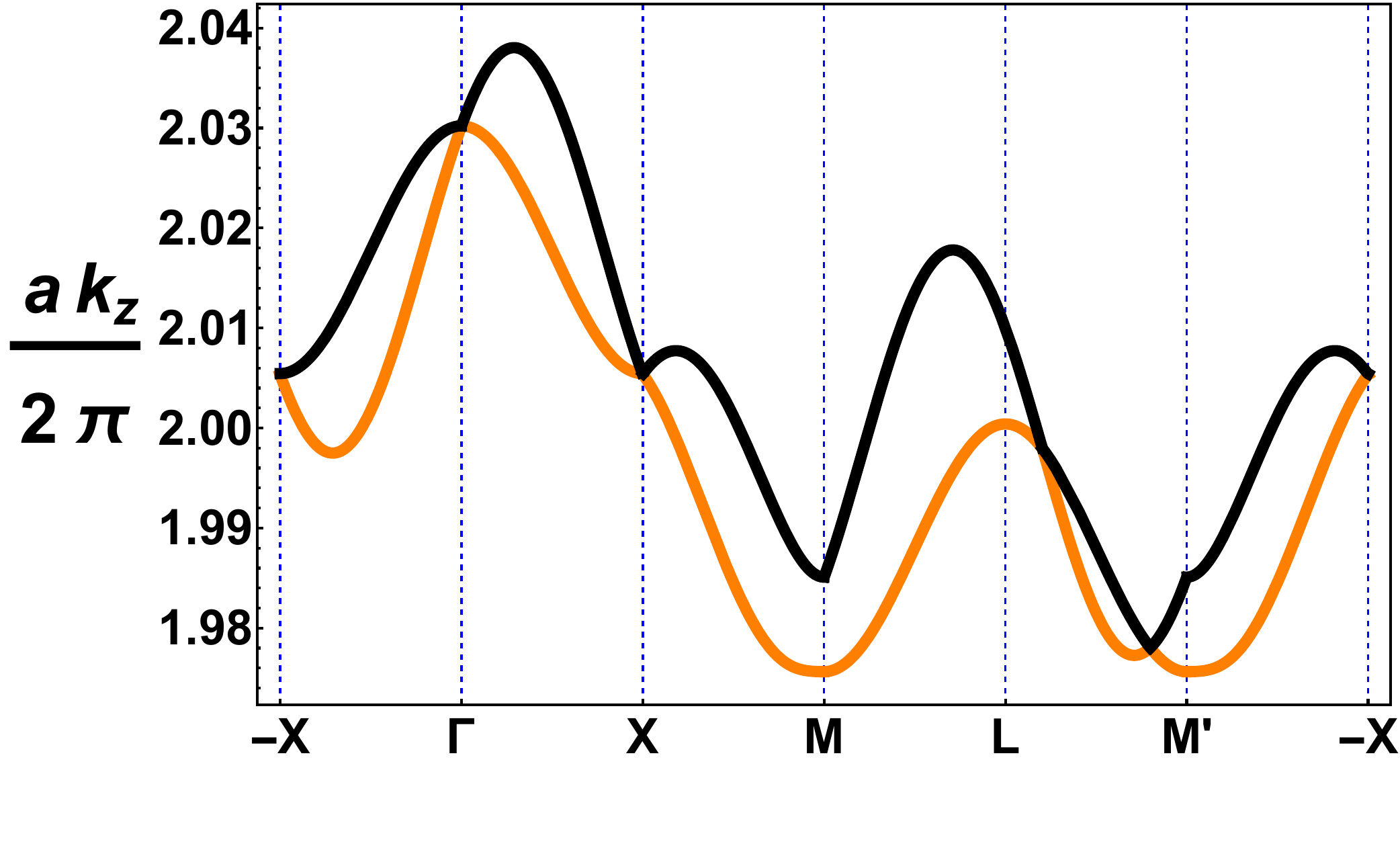}
  \end{minipage}
 \begin{minipage}{0.45\textwidth}
    \includegraphics[width=\textwidth]{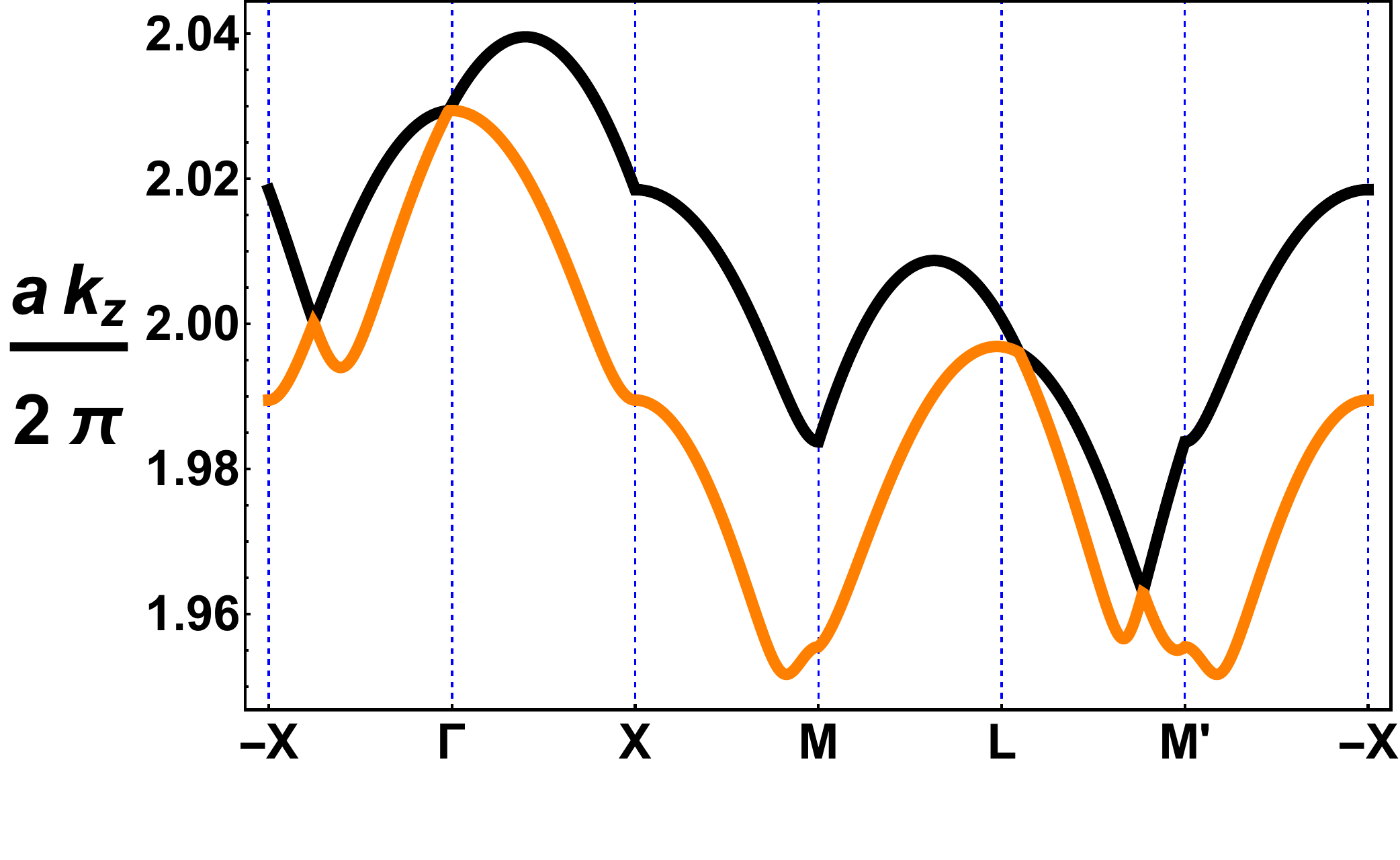}
  \end{minipage}
  \caption{(Color online) Index surfaces along high-symmetry directions of the first Brillouin zone from the paraxial model (top panel) and a frequency-domain plane-wave calculation (lower panel). The relevant high-symmetry points are indicated in Fig.\ \ref{Fig2}.  $ka\simeq 4\pi$, and all other parameters are as in Figs. \ref{Fig2} and \ref{Fig4}.}
  \label{Fig5}
\end{figure}

In Figure \ref{Fig5} we compare the index surfaces of the two
theories, along a high-symmetry path through the Brillouin zone, for
$ka\simeq 4\pi$. This degree of paraxiality places us in a regime towards
the top of the vertical scale of the plots in Figure \ref{Fig4}, in
which there are four Dirac points in the Brillouin zone. As can be
seen in this figure, there is a reasonable correspondence between the
forms of the index surfaces of the two theories. As expected the shapes of the
bands are somewhat different. Nonetheless, given the correspondence
between the numbers and locations of the Dirac points across most degrees
of paraxiality, the topological phase diagrams of the two models will be similar
 to one another. In the regime before the additional Dirac points emerge in the model $H_l$ the 
phase diagram of the numerical simulation will be richer than that of the model.
 
\section{Discussion}

In this paper we have explored the refractive index surfaces of
two-dimensional photonic crystals. We have shown how optical activity
can lead to photonic materials characterized by gapped,
i.e. non-degenerate, index surfaces with non-zero Chern numbers. Our
approach is unusual in that it does not rely on a specific lattice
geometry, but instead on the conical intersections that occur
generically in the index surface of optically anisotropic
materials. This suggests that such topological photonic materials can
be constructed with a range of lattice geometries.

For brevity we have referred to our material as a photonic Chern
insulator. However, as can be seen from Fig.\ \ref{Fig5}, it generally
lacks a band gap at a particular propagation constant, and so is more
accurately described as a photonic semi-metal with topological
bands. In a true Chern insulator the topology of the bulk bands leads,
in the presence of an interface, to the appearance of edge-states in
the gap. For the semi-metal one does expect that a slow spatial
variation of the structure from one topological phase to another will
introduce modes crossing the bulk bands. However, a generic realistic
interface would allow these modes to mix with the bulk bands and
destroy their localization. Given that our proposal is not dependent
on the particular form of lattice geometry, however, it should be
possible to adapt the lattice to achieve a true Chern
insulator. Furthermore, although one does not necessarily expect
conventional edge states, topological semi-metals are known to have
interesting properties in their own
right\ \cite{burkov_topological_2016}.

\begin{acknowledgments}

We acknowledge support from the Irish Research Council award
GOIPG/2015/3570 and Science Foundation Ireland award 15/IACA/3402.

\end{acknowledgments}

\appendix*
\section{Paraxial Hamiltonian}

In this appendix we give the forms of the paraxial Hamiltonian $H_p$
for propagation close to the optic axis, in materials with biaxiality
and anisotropic optical activity. They extend the results in
Ref. \onlinecite{Jfry} to include all terms up to second-order in the
off-axis wavevectors, which is necessary to achieve a non-trivial
topology for the periodic form of $H_p$. These expressions are
obtained by rotating the spatial coordinates so that $z$ lies along an
optic axis, expanding the expressions for $k_{z,\pm}$ to construct a
diagonalized form for the paraxial Hamiltonian in the eigenbasis of
$M$, and transforming the result out of this eigenbasis.
For comparison with Ref.\ \onlinecite{Bry4} we use the scaled
transverse wavevectors $p_{x,y}=k_{x,y}/k=\tilde{k}_{x,y}/\sqrt{\epsilon_2}$,
where $k=k_0\sqrt{\epsilon_2}=\omega\sqrt{\epsilon_2}/c$ is the
wavevector at frequency $\omega$ in an isotropic material of
permittivity $\epsilon_2$. Note however that we use a circular
polarization basis rather than the linear one of
Ref.\ \onlinecite{Bry4}.

Considering first a biaxial dielectric without optical activity,
characterized by principal dielectric constants
$\epsilon_1<\epsilon_2<\epsilon_3$, we find
\begin{equation}
\begin{split}
h_{0,B}&=-A p_x +\frac{1}{4}[2+\epsilon_2 (\alpha-\beta)-12A^2]p_x^2+\\
&+\frac{1}{4}[2+\epsilon_2 (\alpha-\beta)-6A^2]p_y^2.\\
h_{x,B}&=-A p_x +\frac{1}{4}[\epsilon_2 (\alpha-\beta)-12A^2]p_x^2\\
&\quad -\frac{1}{4}\epsilon_2(\alpha+3\beta)p_y^2, \\
h_{y,B}&=-A p_y +\frac{1}{2}[\epsilon_2 (\alpha+\beta)-6A^2]p_x p_y. \\
h_{z,B}&=0.
\end{split}
\end{equation}
Here $\alpha=\epsilon_1^{-1}-\epsilon_2^{-1}$ and $\beta=\epsilon_2^{-1}-\epsilon_3^{-1}$ are measures of the spread of the principal dielectric constants of the biaxial material and $A$ is the biaxial cone semi-angle, defined as $A=\frac{1}{2}\epsilon_2\sqrt{\alpha\beta} $. Sufficiently close to the optic axis direction this Hamiltonian takes a Dirac point form.\\

Optical activity introduces an antisymmetric contribution to the
inverse of the relative permittivity tensor
$\eta_{ik}=(\epsilon^{-1})_{ik}$, corresponding to a contribution
$\epsilon_0 \bm{E}=i \bm{D}\times\bm{g}$ in the constitutive relation,
where $\bm{g}$ is the optical activity vector
\cite{Bry1,landau_electrodynamics_2008}. In the case of the Faraday
effect the optical activity vector is proportional to the applied
field, $g_i=\gamma_{ij} H_j$, and independent of the wavevector. In
the paraxial Hamiltonian this introduces additional terms in $h_{0}$,
describing the field's effect on the overall dispersion relation, as
well as terms in $h_{3}$, describing the anisotropic splitting between
the circular polarization states. The expressions for these components
are
\begin{equation}
\begin{split}
h_{0,F}&=h_{0,B}-\frac{3}{8}\epsilon_2^2\big[g'^2_3+2g'_3(g'_1 ~ p_x+g'_2~p_y) +\\
&+(g'^2_1-g'^2_3)p_x^2 +2g'_1g'_2~p_xp_y + (g'^2_2-g'^2_3)p_y^2\big].
\end{split}
\end{equation}

and\\

\begin{equation}
\begin{split}h_{z,F}&=\frac{1}{2}\epsilon_2\bigg(g'_3 +(g'_1+3Ag'_3)p_x +g_2'p_y \\
&\qquad\quad  -\frac{1}{2} \{g'_3[1+\frac{3}{2}\epsilon_2(\alpha-\beta)]-6Ag'_1\}p_x^2 \\ &\qquad\quad+3Ag'_2 p_xp_y-\frac{1}{2}\{g'_3[1+\frac{3}{2}\epsilon_2(\alpha-\beta)]\}p_y^2\bigg). \\
\end{split}
\end{equation}
Here $\bm{g'}$ is the optical activity vector in the rotated basis,
where $z$ lies along the optic axis. The relation to the optical
activity vector in the principal axes coordinate system (where
$\epsilon$ is diagonal) is $\bm{g'}=R\bm{g}$ with
\begin{equation}
R=\left(\begin{array}{ccc}
\sqrt{\frac{\beta}{\alpha+\beta}} & 0 & -\sqrt{\frac{\alpha}{\alpha+\beta}} \\
0 & 1 & 0\\
\sqrt{\frac{\alpha}{\alpha+\beta}} & 0 &\sqrt{\frac{\beta}{\alpha+\beta}}
\end{array}\right).\label{eq:coords}
\end{equation}

For completeness we note that, in the case of a chiral medium, the
optical activity vector is related to the wavevector direction,
$g_i=\Gamma_{ij}\hat{k}_j$, by a symmetric tensor $\Gamma_{ij}$
\cite{Bry4,landau_electrodynamics_2008}. This modifies the components
$h_0$ and $h_3$ as follows:\begin{equation}
\begin{split}
h_{0,C}&=h_{0,B} -\frac{3}{8}\epsilon_2^2
\big\{ \Gamma'^2_{33}+4\Gamma'_{13} \Gamma'_{33} p_x + 4\Gamma'_{23}\Gamma'_{33}p_y +\\
&+[4\Gamma'^2_{13}  +2\Gamma'_{33}(\Gamma'_{11}-\Gamma'_{33})]p_x^2+\\
&+4(\Gamma'_{12} \Gamma'_{33} + 2\Gamma'_{13}\Gamma'_{23})p_xp_y+\\
&+[4\Gamma'^2_{23}+2\Gamma'_{33}(\Gamma'_{22}-\Gamma'_{33})]p_y^2\big\}.
\end{split}
\end{equation}
and

\begin{equation}
  \begin{split}
    h_{z,C}&=\frac{1}{2}\epsilon_2\bigg(\Gamma'_{33}+(2\Gamma'_{13}+3A\Gamma'_{33}) ~p_x+2\Gamma'_{23}~p_y\\&\qquad\qquad+\{\Gamma'_{11}+6A\Gamma'_{13}-[1+\frac{3}{4}\epsilon_2(\alpha-\beta)]\Gamma'_{33}\}p_x^2\\
&\qquad\qquad+ (2\Gamma'_{12}+6A\Gamma'_{23})~p_xp_y  \\
&\qquad\qquad+\{\Gamma'_{22} -[1+\frac{3}{4}\epsilon_2(\alpha-\beta)]\Gamma'_{33}\}p_y^2\bigg). \\
\end{split}
\end{equation} where the tensor $\Gamma'$ is related to the corresponding tensor in the principal axes basis by $\Gamma'=R\Gamma R^{-1}$.


%

\end{document}